\documentclass[11pt]{article}

\usepackage{geometry}  
\usepackage[T1]{fontenc}  
\usepackage[utf8x]{inputenc}  
\usepackage{lineno}  
\usepackage{fancyhdr}  
\usepackage{enumitem}  
\usepackage{hyperref}  
\usepackage{titling}  
\usepackage[numbers,sort&compress,round]{natbib}  
\usepackage{mathtools}  
\usepackage{titlesec}  
\usepackage{lastpage}  

\usepackage[english]{babel}  
\usepackage{amsmath}  
\usepackage{amsfonts}  
\usepackage{amssymb}  
\usepackage{wasysym}  
\usepackage{bbm}  
\usepackage{array}  
\usepackage{xr}  
\usepackage{verbatim} 
\usepackage{float} 
\usepackage{graphicx}
\usepackage{caption}
\usepackage{subcaption}
\usepackage{multirow}

\hypersetup{%
  pdfborderstyle={/S/U/W 1}
}

\titleformat{name=\section}{\normalfont\Large\bfseries}{}{0pt}{}
\titleformat{name=\subsection}{\normalfont\large\bfseries}{}{0pt}{}
\titleformat{name=\subsubsection}{\normalfont\normalsize\bfseries}{}{0pt}{}

\newcommand{\email}[1]{\href{mailto:{#1}}{{#1}}}

\newcommand{\keywords}[1]{\textbf{Keywords}: {#1}}

\newcommand{\optincludegraphics}[2][]{}

\newcommand{\optinput}[1]{}

\newtagform{brackets}{[}{]}
\usetagform{brackets}

\title{Game of Learning Bloch Equation Simulations for MR Fingerprinting}
\begin{document}

\begin{titlepage}
{\noindent\LARGE\bf \thetitle}

\bigskip

\begin{flushleft}\large
	Mingrui~Yang\textsuperscript{1,{*}},
	Yun~Jiang\textsuperscript{2},
	Dan~Ma\textsuperscript{2},
	Bhairav~B.~Mehta\textsuperscript{2},
        Mark~A.~Griswold\textsuperscript{2,3}
\end{flushleft}

\bigskip

\noindent
1. Department of Biomedical Engineering, Cleveland Clinic, Cleveland,
Ohio, USA\\
2. Department of Radiology, Case Western Reserve University, Cleveland,
Ohio, USA\\
3. Department of Biomedical Engineering, Case Western Reserve University,
Cleveland, Ohio, USA
\bigskip


\noindent{*} Correspondence to:


Mingrui~Yang	\\
\indent Department of Biomedical Engineering\\
\indent Lerner Research Institute\\
\indent Cleveland Clinic\\
\indent 9500 Euclid Ave\\
\indent Cleveland, OH 44106, USA									\\
\indent Email: \email{yangm@ccf.org}\\





\end{titlepage}

\pagebreak

\begin{abstract}

\textbf{Purpose}: This work proposes a novel approach to efficiently generate MR
fingerprints for MR fingerprinting (MRF) problems based on the unsupervised
deep learning model generative adversarial networks (GAN).

\textbf{Methods}: The GAN model is adopted and modified for better
convergence and performance, resulting in an MRF specific model named
GAN-MRF. The GAN-MRF model is trained, validated, and tested using
different MRF fingerprints simulated from the Bloch equations with
certain MRF sequence. The performance and robustness of the model are further tested
by using in vivo data collected on a 3 Tesla scanner from a healthy
volunteer together with MRF dictionaries with different sizes. $T_1$,
$T_2$ maps are generated and compared quantitatively.

\textbf{Results}: The validation and testing curves for the GAN-MRF
model show no evidence of high bias or high variance problems. The
sample MRF fingerprints generated from the trained GAN-MRF model agree
well with the benchmark fingerprints simulated from the Bloch
equations. The in vivo $T_1$, $T_2$ maps generated from the GAN-MRF
fingerprints are in good agreement with those generated from the Bloch
simulated fingerprints, showing good performance and robustness of the
proposed GAN-MRF model. Moreover, the MRF dictionary generation time
is reduced from hours to sub-second for the testing dictionary.

\textbf{Conclusion}: The GAN-MRF model enables a fast and accurate
generation of the MRF fingerprints. It significantly reduces the MRF
dictionary generation process and opens the door for real-time
applications and sequence optimization problems.

\end{abstract}

\bigskip

\pagebreak

\section{Introduction}

Magnetic resonance (MR) imaging has been a successful diagnostic
imaging modality due to its ability to characterize a wide range of
underlying tissue parameters. However, traditional MR images are
generally qualitative, and can vary from scan to scan, leading to a
variability in interpretation and limitation in objective
evaluation. The aims of quantitative MR imaging is to eliminate this
variability and limitation, and provide additional pathological
information for diagnosis in a quantitative and deterministic
manner. Conventional quantitative MR imaging methods for MR parameter
(e.g. $T_1$, $T_2$, $T_2^*$) mapping has had many successful
applications in both research and clinical
settings~\cite{MRM:MRM1910110308,Usman782,Payne738,VANHEESWIJK20121231,ILES20081574,JMRI:JMRI24584,JMRI:JMRI22775,JON:JON199443146,MRM:MRM21923,Souza2013}. Most
of these methods, however, can only evaluate one parameter of interest
at a time, resulting in a prolonged scan time when multiple
acquisitions have to be repeated to obtain different parameter maps
for clinical applications. 

Magnetic resonance fingerprinting (MRF)~\cite{ma13} is a newly
developed quantitative magnetic resonance imaging method that may overcome some previous limitations of quantitative MR imaging~\cite{Cloos2015,Gomez2016,Chen2016a,Cohen2016,Doneva2016,Cline2016,Zhang2016,Chen2016b,Zhao2016,Pahwa2016,7590737}. Unlike conventional quantitative MR imaging methods, MRF is able to efficiently obtain multiple tissue property maps simultaneously within one single scan. The collected raw data from a scanner is first reconstructed using e.g. nonuniform fast Fourier transform~\cite{1166689}. The time dimension of each reconstructed voxel is then matched against a pre-calculated MRF dictionary using Bloch simulations, which is one of the key components of MRF. Depending on the tissue properties of interest, the dictionary can be calculated for different MRF sequences, such as the balanced steady-state free precession
(bSSFP)~\cite{ma13} sequence, the fast imaging steady-state precession
(FISP)~\cite{MRM:MRM25559} sequence, or the MRF-X~\cite{Hamilton2015}
sequence. The size of the MRF dictionaries generated changes with the MRF sequence chosen and the step size used for certain tissue properties. It can be prohibitively large if complex sequences considering multiple tissue properties or fine step size for tissue properties are used. This can make the pattern matching stage of MRF significantly slowed down, or even worse, completely paralyzed due to lack of computer memory. Efforts have been taken to speed up the MRF pattern matching process~\cite{6851901,MRM:MRM25439}. These methods,
however, still rely on a full sized MRF dictionary, and therefore,
cannot resolve the memory consumption
problem. Yang~et~al.~\cite{MRM:MRM26867} proposed to use the randomized singular value decomposition together with polynomial fitting methods to significantly reduce the memory overhead and speed up the pattern matching step of MRF problems. Nevertheless, none of these methods has considered the time needed to generate MRF dictionaries. In fact, the time required for generating these dictionaries varies, but can be prohibitively long, especially when many factors are included into the calculation. For example, a slice profile corrected FISP dictionary requires the simulation of multiple spin evolutions which are then summed for each time frame to average out the effect of off resonance. Some dictionary calculations that involve exchange and other complicated physics can take days or even weeks to calculate~\cite{Hamilton2015,doi:10.1002/mrm.27040}. 


In this paper, we present a new approach to create MRF dictionaries
with a significantly reduced time cost based on the recent development
in the deep learning community. Specifically, we modify one of the
most interesting unsupervised models, the
generative adversarial networks (GAN)~\cite{Goodfellow2014}, into a
semi-supervised model for our purpose, fed with tissue parameter
combinations and sequence parameters. Given the trained GAN-MRF model,
the problem of generating MRF dictionaries through the
complicated Bloch equation simulations is transformed into easy
matrix multiplications followed by some simple nonlinear activation
functions. This transformation can significantly reduce the time needed to generate MRF
dictionaries, which makes it possible to generate dictionaries with
tissue properties of interest on-the-fly. We believe that this will
open the door to the rapid calculation of dictionaries with more
complex physics as well. In vivo 3T brain scan data are used to
evaluate the quality of the MRF dictionaries generated.

\section{Theory}


In this section, we present the details of the GAN model and its
limitations. We then describe in detail our modified GAN-MRF model to
possibly address these limitations.
\begin{figure}[ht!]
  \includegraphics[scale=.4]{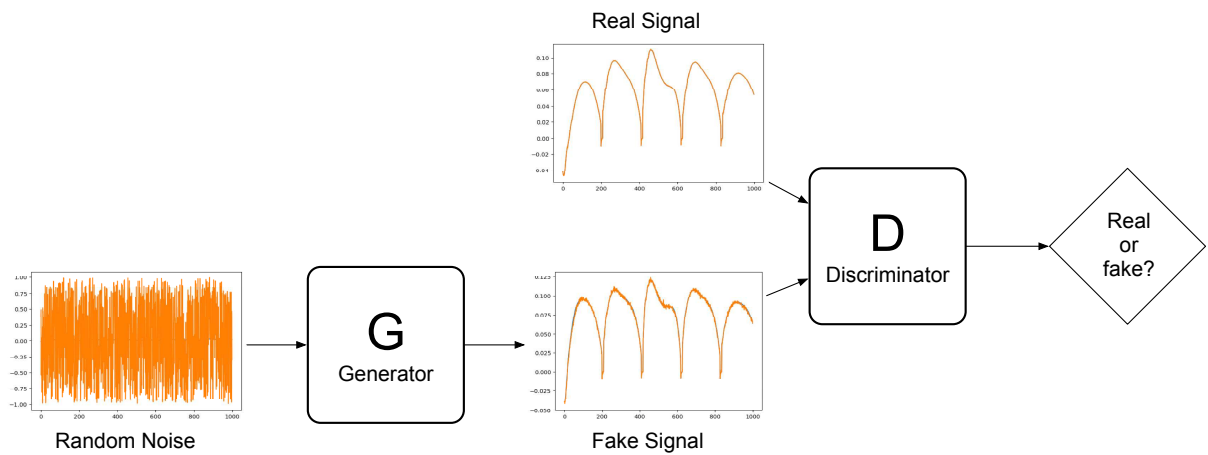}
  \caption{Generative Adversarial Network}
  \label{fig:gan}
\end{figure}
GAN is a newly developed unsupervised machine learning model, which has been
vigorously studied in the past few
years~\cite{NIPS2015_5773,NIPS2014_5423,DBLP:journals/corr/RadfordMC15,DBLP:journals/corr/SalimansGZCRC16,DBLP:journals/corr/ZhaoML16,DBLP:journals/corr/ReedAYLSL16,Odena16,DBLP:journals/corr/ChenDHSSA16,DBLP:journals/corr/IsolaZZE16,doi:10.1002/mrm.27627}. It
basically contains a pair of networks competing with each other: a
generative network (generator) and a discriminative network (discriminator). The
generator is trained to synthesize data samples that mimic the real
data from pure random noise to
fool the discriminator;
while the discriminator is trained to distinguish the real data
samples from the synthesized samples generated by the generator, as illustrated in Fig.~\ref{fig:gan}. They act as two players playing a minimax game and
achieving the Nash equilibrium eventually. 

Let us consider the generator $G_{\theta_g}(z): \mathbb{R}^m \rightarrow
\mathbb{R}^n$ as a function mapping from a fixed prior noise
distribution $p(z)$ to the data space, and the discriminator
$D_{\theta_d}(x): \mathbb{R}^n \rightarrow [0,1]$ as a function mapping
from the data space to a probability, where $\theta_g$ and $\theta_d$
are the parameters to estimate for the generator and discriminator
networks respectively. Then the minimax objective function can be
written as

\begin{align}
  \min_{\theta_g}\max_{\theta_d}\left[\mathbb{E}_{x\sim p_{data}}\log
  D_{\theta_d}(x) + \mathbb{E}_{z\sim
  p(z)}\log(1-D_{\theta_d}(G_{\theta_g}(z)))\right],  \label{eq:minimax}
\end{align}
where $x$ is drawn
from the real data population $p_{data}$, and $z$ is drawn from the
prior noise distribution $p(z)$. Notice that this is a non-convex
optimization problem. The convergence to the global optima cannot be
guaranteed. A typical approach to solve this minimax optimization
problem is to alternate between

\begin{align}
  \max_{\theta_d}\left[\mathbb{E}_{x\sim p_{data}}\log
  D_{\theta_d}(x) + \mathbb{E}_{z\sim p(z)}\log(1-D_{\theta_d}(G_{\theta_g}(z)))\right]
\end{align}
and

\begin{align}
  \min_{\theta_g}\mathbb{E}_{z\sim
  p(z)}\log(1-D_{\theta_d}(G_{\theta_g}(z)))\label{eq:g-ascent}.
\end{align}
Note that solving [\ref{eq:g-ascent}] with the gradient descent algorithm is
not efficient, since when the discriminator is not good, the gradient is
small, which cannot provide sufficient information for the generator
to evolve. Therefore, in practice, [\ref{eq:g-ascent}] is often transformed
to an equivalent form

\begin{align}
  \max_{\theta_g}\mathbb{E}_{z\sim p(z)}\log D_{\theta_d}(G_{\theta_g}(z))
\end{align}
for faster convergence.

The GAN model, although exciting, has several known issues. First of
all, as mentioned above, the minimax problem [\ref{eq:minimax}] is a
non-convex optimization problem, whose convergence to a global optima
is not guaranteed. In other words, the problem may only converge to a
local optima, or even worse, get stuck at a saddle point, which is
neither a local maxima nor a local minima. Second, the model may collapse into a single mode
so that the generator learns a pattern to generate a unique good looking fake
data to fool the discriminator over and over again. Moreover, even if the
GAN model does not collapse, the trained  generator can only generate
a limited number of distinct samples.

Our goal here is that, after the model is trained, we would like the
generator to be able to synthesize a large variety of MR fingerprints
corresponding to a wide range of tissue property and sequence
parameter combinations. When used without modification, the
limitations of the GAN model mentioned above dominate its performance,
rendering it unable to fulfill our purpose. Therefore, we need to
modify the GAN model for our purpose. It has been shown in the
literature that the problems mentioned above can be partially solved
by adding conditional information and regularization terms into the
model~\cite{DBLP:journals/corr/IsolaZZE16,Odena16}. We follow these ideas to modify the GAN model and write our GAN-MRF model as

\begin{align}
  \min_{\theta_g}\max_{\theta_d}\left[\mathbb{E}_{x\sim p_{data}}\log
  D_{\theta_d}(x|y) + \mathbb{E}_{z\sim
  p(z)}\log(1-D_{\theta_d}(G_{\theta_g}(z|y)|y)) +
  \lambda\mathbb{E}_{x\sim p_{data}, z\sim
  p(z)}\|x-G_{\theta_g}(z|y)\|_1\right], \label{eq:minimax_mod} 
\end{align}
where $x$ is drawn from the training fingerprints simulated from Bloch
equations, $y$ is the control variable concatenating the corresponding
sequence parameters and tissue parameter combinations, $z$ is drawn
from the normal distribution $\mathcal{N}(0,1)$, and $\lambda$ is a
hyperparameter controlling the regularization term. The conditional variable y can be a combination of, for instance, flip angle and repetition time, which are fed into the model in addition to the simulated fingerprints to better regulate the behavior of the model. $\ell_1$ regularization is used since it is known to be more robust than e.g. the Euclidean distance regularization to noise and outliers, which is important for MR fingerprints generation. A small perturbation in an MR fingerprint can lead to completely different interpretation of the underlying tissue properties. The choice of the hyperparameter $\lambda$ can be determined through a model validation process as explained in details in the Method section. The mini-max problem~\eqref{eq:minimax_mod} is again a non-convex optimization problem, which can be solved by alternating between

\begin{align}
  \max_{\theta_d}\left[\mathbb{E}_{x\sim p_{data}}\log
  D_{\theta_d}(x|y) + \mathbb{E}_{z\sim
  p(z)}\log(1-D_{\theta_d}(G_{\theta_g}(z|y)|y))\right]
  \label{eq:D_loss}
\end{align}
and

\begin{align}
  \max_{\theta_g}\left[\mathbb{E}_{z\sim p(z)}\log
  D_{\theta_d}(G_{\theta_g}(z|y)|y) - \lambda\mathbb{E}_{x\sim
  p_{data}, z\sim p(z)}\|x-G_{\theta_g}(z|y)\|_1\right].
  \label{eq:G_loss}
\end{align}

\section{Methods}



\subsection{Data Generation}
\begin{figure}[h!]
  \centering
  \begin{subfigure}{0.4\textwidth}
    \includegraphics[width=\textwidth]{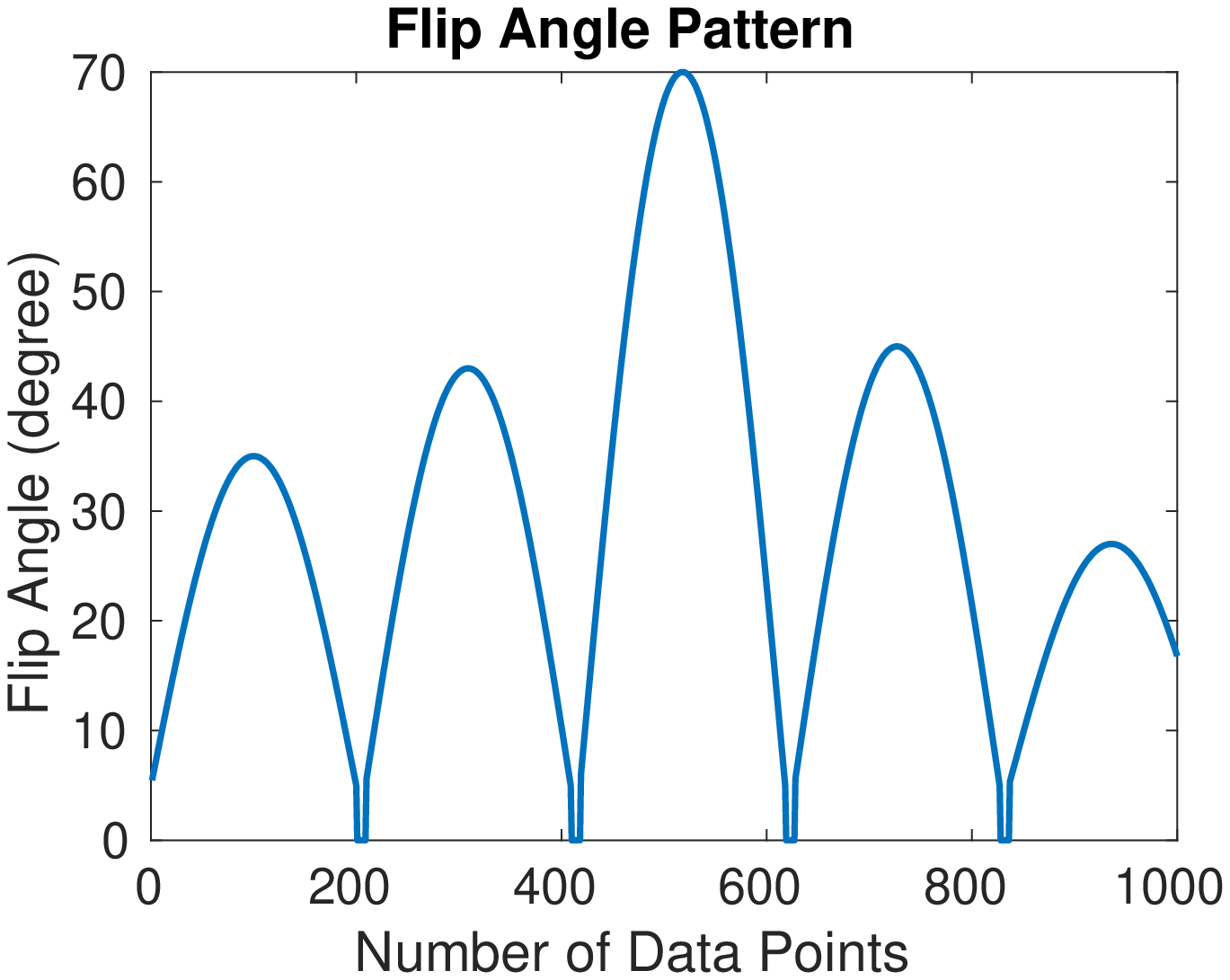}
    \caption{}
    \label{fig:fa}
  \end{subfigure}
  \begin{subfigure}{0.4\textwidth}
    \includegraphics[width=\textwidth]{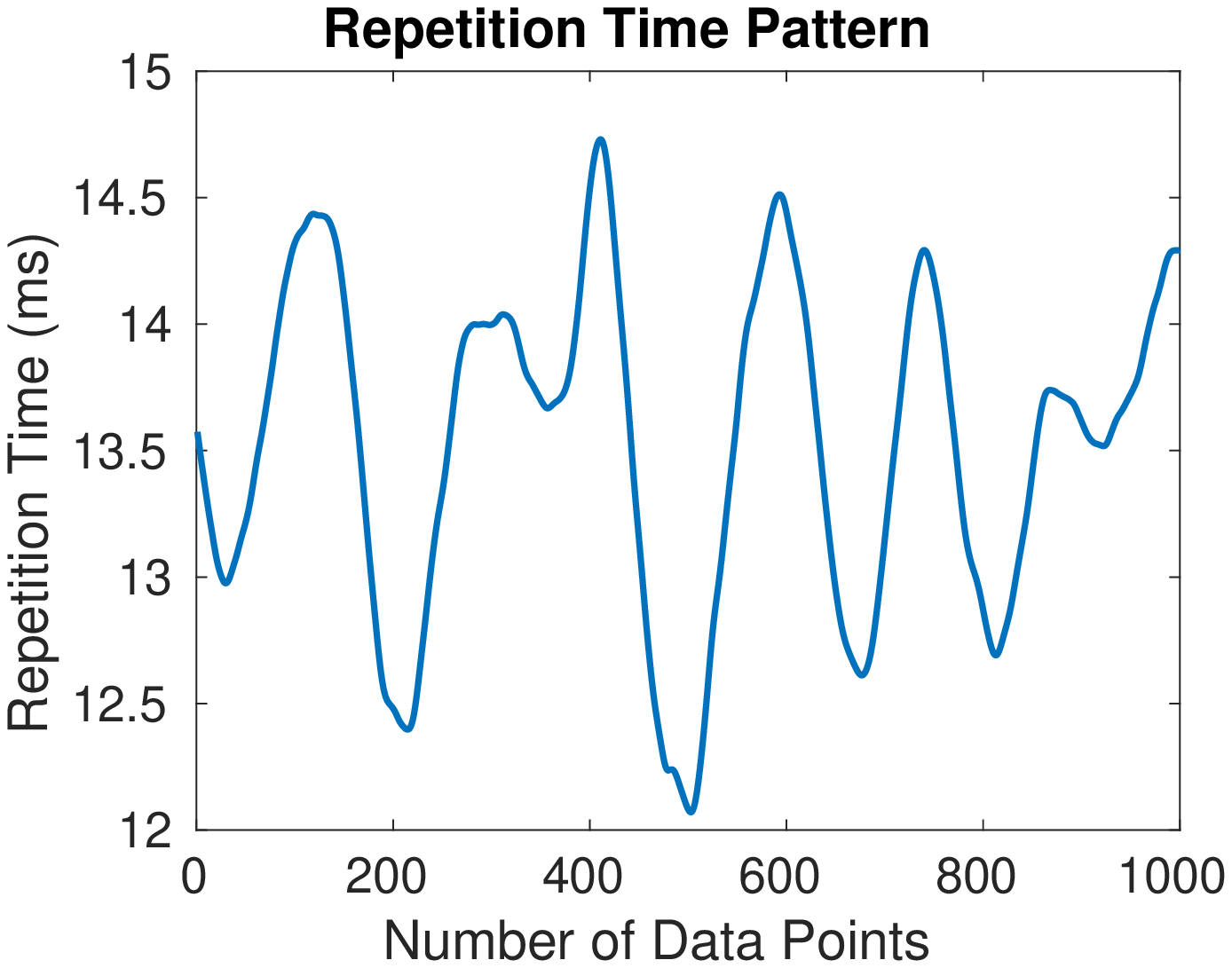}
    \caption{}
    \label{fig:tr}
  \end{subfigure}
  \caption{The patterns of the flip angle and repetition time used in
    the FISP sequence.}
  \label{fig:fa+tr}
\end{figure}

The data we used to train, validate, and test the GAN-MRF model was an
MRF dictionary generated from Bloch equation simulations using a FISP
sequence with slice profile correction~\cite{}. The
$T_1$ values chosen for the simulations ranged from 10ms to
2950ms. The $T_2$ values ($\le T_1$) range from 2ms to 500ms. They lead to a total
tissue parameter combinations of 5970. The details of the
ranges and step sizes of $T_1$ and $T_2$ values are listed in
Table~\ref{tab:dict_std}. The patterns of the flip angles and
repetition time are shown in Fig.~\ref{fig:fa+tr}, with the flip
angles ranging from 5 degrees to 70 degrees, and the repetition time
ranging from 12.07ms to 14.73ms, resulting in a total of 1000 time
frames. Note that the 1000 time frames are considered as 1000 features
and the 5970 $T_1$, $T_2$ combinations are considered as examples. The
dataset was further divided into three parts including training
data, validation data, and test data, so that each part contains 60\%,
20\%, and 20\% of the total 5970 dictionary atoms respectively. The
training, validation, and test sets were then normalized separately to
avoid interference to the validation and test results from the
training data.

\begin{table}[h!]
\begin{center}
  \begin{tabular}{c|c|c}
    \hline
    & Range & Step Size \\
    \hline
    \multirow{5}{*}{$T_1$} & $[10, 85]$ & 5 \\
         & $[90, 990]$ & 10 \\
         & $[1000,1480]$ & 20 \\
         & $[1500,2000]$ & 50 \\
         & $[2050,2950]$ & 100 \\
    \hline
    \multirow{4}{*}{$T_2$} & $[2,8]$ & 2 \\
         & $[10,145]$ & 5 \\
         & $[150,190]$ & 10 \\
         & $[200,500]$ & 50 \\
    \hline
  \end{tabular}
\end{center}
\caption{Ranges and step sizes of $T_1$, $T_2$ values. All in
  milliseconds (ms).}
\label{tab:dict_std}
\end{table}

\subsection{Model Specifics}
\begin{figure}
  \centering
  \includegraphics[scale=.4]{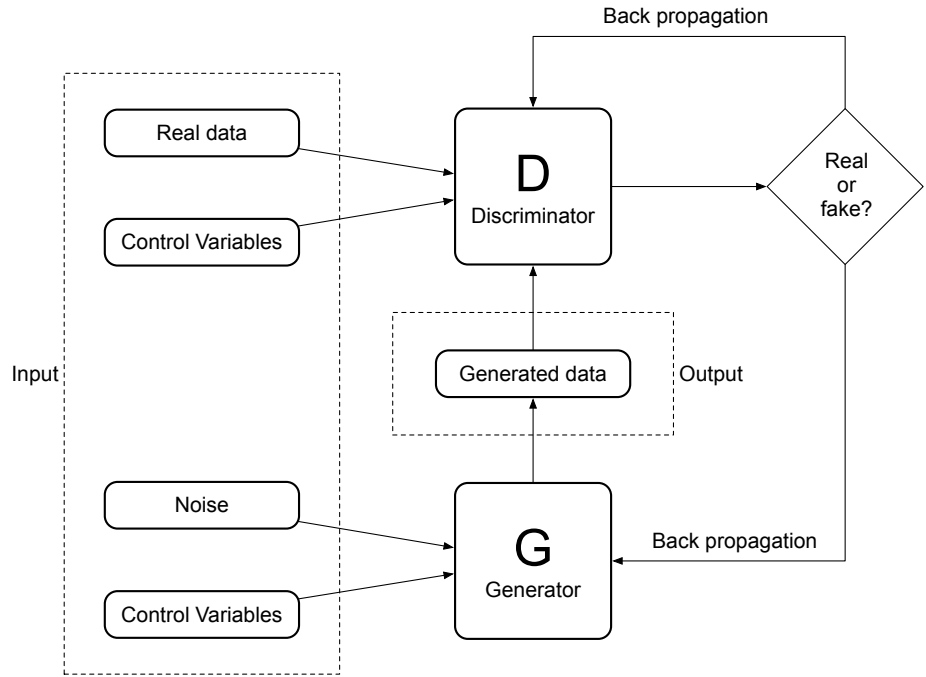}
  \caption{Flow chart for GAN-MRF}
  \label{fig:mrfgan}
\end{figure}

The input layer of the discriminative network took MR fingerprints,
( either simulated from the Bloch equations with the FISP sequence, or
synthesized by the generative network,) together with the
corresponding $T_1$ and $T_2$ combinations. They were then passed
through 3 hidden layers, each consisting of 128 neurons, followed by a
rectified linear unit (ReLU). The output layer of the discriminative
network outputed a probability of the input fingerprint being Bloch
equations simulated by applying a sigmoid function as the activation function. The input layer of
the generative network took pure random noise signals, together with
the desired $T_1$, $T_2$ combinations and sequence parameters such as
flip angels and repetition time. Similar to the discriminative
network, they were then passed through 3 hidden layers, each containing
128 neurons followed by a ReLU. The output layer of the generative
network synthesized the corresponding MR fingerprints by utilizing a
hyperbolic tangent function as the activation function.In summary, the discriminative network
took training fingerprints and the associated sequence parameters and
tissue property combinations to improve its performance in
distinguishing real and synthesized fingerprints. The generative
network only needed to know the input of sequence parameters and tissue
parameter combinations of interest, and outputs the fingerprints that
mimicking the real ones. A block diagram summarizing the GAN-MRF
architecture is shown in Fig.~\ref{fig:mrfgan}. 

\subsection{Model Training, Validating, and Testing}
The training, validation, and testing of the model were realized by
using the Python deep learning framework
Tensorflow~\cite{DBLP:journals/corr/AbadiABBCCCDDDG16} on a Ubuntu
system with an Intel Xeon 2.6GHz CPU (Intel Corporation, Santa Clara,
CA). 

The discriminative network was trained by applying the stochastic gradient
descent method  to small batches of the training and synthesized data
with the minibatch size equal to 30, using the Adam
optimizer~\cite{DBLP:journals/corr/KingmaB14} with a learning rate of
$10^{-5}$. The generative network was trained in the same fashion, but
on small batches of random noise and training data. The loss functions
for training the discriminative network and the generative network were modified
versions of Eq.~\eqref{eq:D_loss} and Eq.~\eqref{eq:G_loss}, where the
maximization problems were converted into minimization problems by
negating both \eqref{eq:D_loss} and \eqref{eq:G_loss}.

The hyperparameter $\lambda$ controlling model regularization in
Eq.~\eqref{eq:G_loss} was determined by running a model validation
on the validation set. First, the root mean square errors between the
training data and the synthesized data from the trained GAN-MRF model
were calculated and plotted against difference choices of the values of
the hyperparameter $\lambda$. Then, the root mean square errors
between the validation data and the synthesized data generated from
the trained model using the validation control variables against
different choices of $\lambda$ were also computed and plotted. The
$\lambda$ value with the smallest validation root mean square error
and fast convergence rate was chosen to be the hyperparameter value
used in the loss function \eqref{eq:G_loss} for the generative network.

After the choice of the hyperparameter $\lambda$ was determined, the
trained GAN-MRF model was further tested on the test set independent of
the training and validation sets for potential model underfitting or
overfitting problems. Specifically, after each iteration, we calculated
the root mean square errors between the synthesized fingerprints from
the GAN-MRF model with the updated model parameters and the benchmark
fingerprints from the training and test sets respectively. The root
mean square errors were then plotted against the number of iterations
to examine for indications of poor convergence, high bias or high variance problems. 

\subsection{In Vivo Test}
To test the performance of our model on \textit{in vivo} dataset, a
GAN-MRF dictionary was generated using the trained generative network
with the same control variables used for the training, validation and
test sets, resulting in a dictionary of size $1000\times5970$. The \textit{in
vivo} brain scan of a healthy volunteer was obtained on a Siemens Skyra
3T scanner (Siemens Healthcare, Erlangen, Germany) with a 20-channel
head receiver coil array. The informed consent was obtained before the
scan. All the experiments were approved by our institutional review
board. The sequence used for the scan was the MRF-FISP sequence with the same sequence parameters and spiral sampling trajectory as previously reported~\cite{MRM:MRM25559} with an acceleration factor of 48 (one out of 48
spiral interleaves per repetition of MRF-FISP acquisition), a matrix size
of 256 × 256, and a FOV of 30 × 30cm$^2$. The collected spiral data from
each coil were reconstructed using the non-uniform fast Fourier transform with an independently
measured spiral trajectory for gradient imperfection
correction~\cite{1166689}. Reconstructed images from all individual coils were then
combined and compensated for coil sensitivity variation. $T_1$, $T_2$ maps
were created by applying the standard MRF pattern matching
algorithm between the reconstructed images and the GAN-MRF
dictionary. The generated $T_1$, $T_2$ maps were then compared to the
benchmark maps generated from the simulated MRF-FISP dictionary to
compute the difference maps and the relative root mean square errors. 

We further tested the scalability of our GAN-MRF model by first
training the GAN- MRF model on a small training set simulated from
Bloch equations. The trained model was then used to synthesize a
much larger MRF dictionary with finer $T_1$ and $T_2$ step sizes. The
synthesized dictionary was then used, together with the \textit{in vivo} data,
to perform the MRF pattern matching to obtain the $T_1$, $T_2$
maps, which were compared against the maps obtained from the MRF-FISP
dictionary by Bloch equation simulations with the same finer $T_1$,
$T_2$ step sizes. More specifically, the coarse MRF-FISP dictionary
(i.e. the training set) contained 1000 time frames and 297
tissue parameter combinations with the same $T_1$, $T_2$ ranges as in Table~\ref{tab:dict_std}. We
then generated a much finer GAN-MRF dictionary containing 106160 tissue property combinations using
the trained model with an input of the refined $T_1$, $T_2$
combinations. $T_1$ and $T_2$ maps for the \textit{in vivo} data were obtained
from the synthesized fine GAN-MRF dictionary and
compared against the ones generated from the MRF-FISP dictionary
simulated directly from the Bloch equations with the refined $T_1$,
$T_2$ combinations. The detailed step sizes of the coarse and fine
$T_1$ and $T_2$ combinations are listed in Table~\ref{tab:dict_coarse_fine}.

\begin{table}[h!]
\begin{center}
  \begin{tabular}{c|c|c|c|c}
    \hline
    & \multicolumn{2}{c|}{Coarse} & \multicolumn{2}{c}{Fine} \\
    \hline
    & Range & Step Size & Range & Step Size \\
    \hline
    \multirow{4}{*}{$T_1$} & $[50, 100]$ & 50 & $[2, 100]$ & 2 \\
         & $[200, 1000]$ & 100 & $[105, 1000]$ & 5 \\
         & $[1200,2000]$ & 200 & $[1010,2000]$ & 10 \\
         & $[2500,3000]$ & 500 & $[2025,3000]$ & 25 \\
    \hline
    \multirow{3}{*}{$T_2$} & $[10,100]$ & 10 & $[1,200]$ & 1 \\
         & $[120,200]$ & 20  & $[202,500]$ & 2 \\
         & $[300,500]$ & 100  &  & \\
    \hline
  \end{tabular}
\end{center}
\caption{Ranges and step sizes of $T_1$, $T_2$ values. All in
  milliseconds (ms).}
\label{tab:dict_coarse_fine}
\end{table}

\section{Results}

\begin{figure}[h!]
  \centering
  \begin{subfigure}{0.45\textwidth}
    \includegraphics[width=\textwidth]{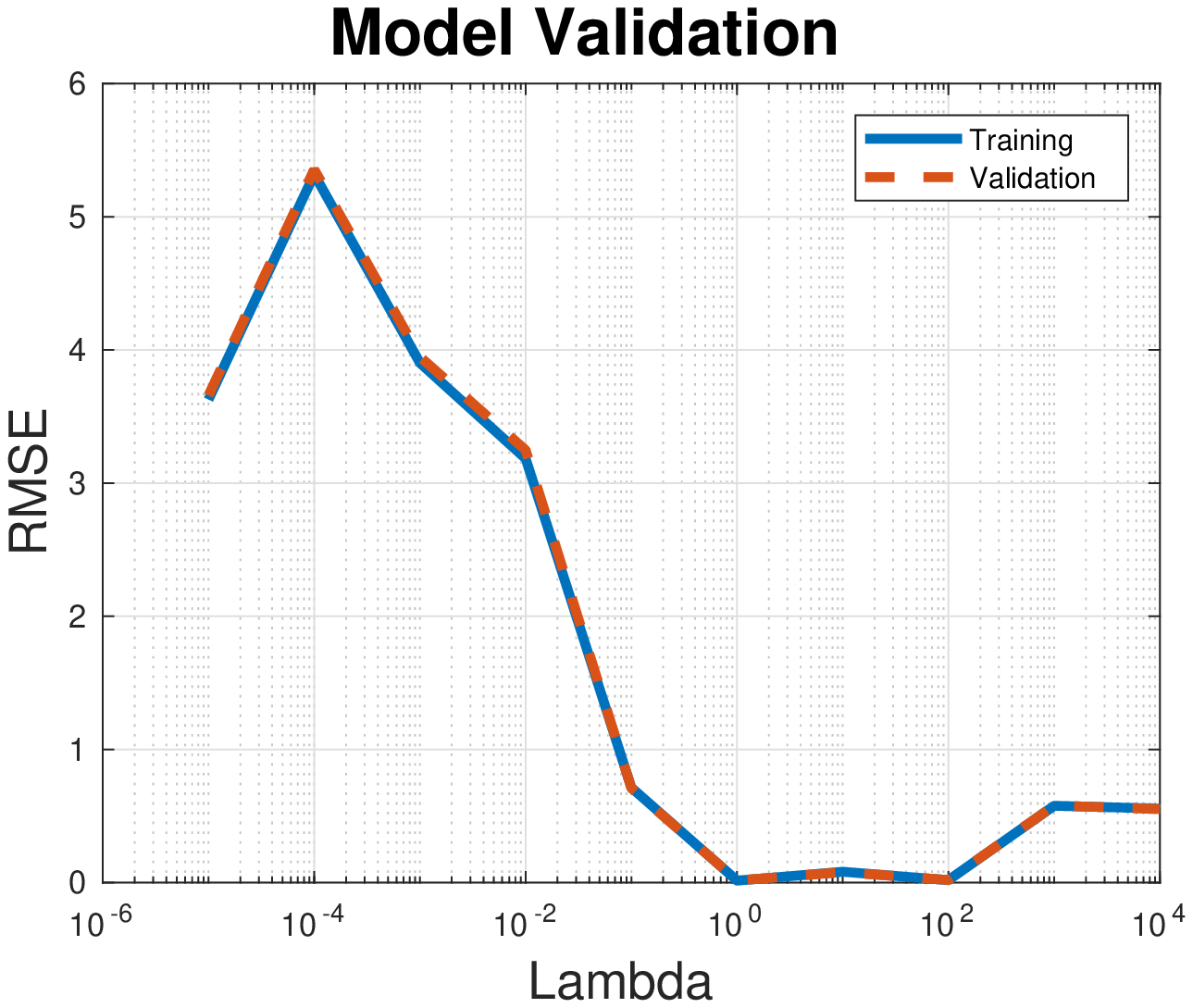}
    \caption{}
    \label{fig:validation}
  \end{subfigure}
  \begin{subfigure}{0.45\textwidth}
    \includegraphics[width=\textwidth]{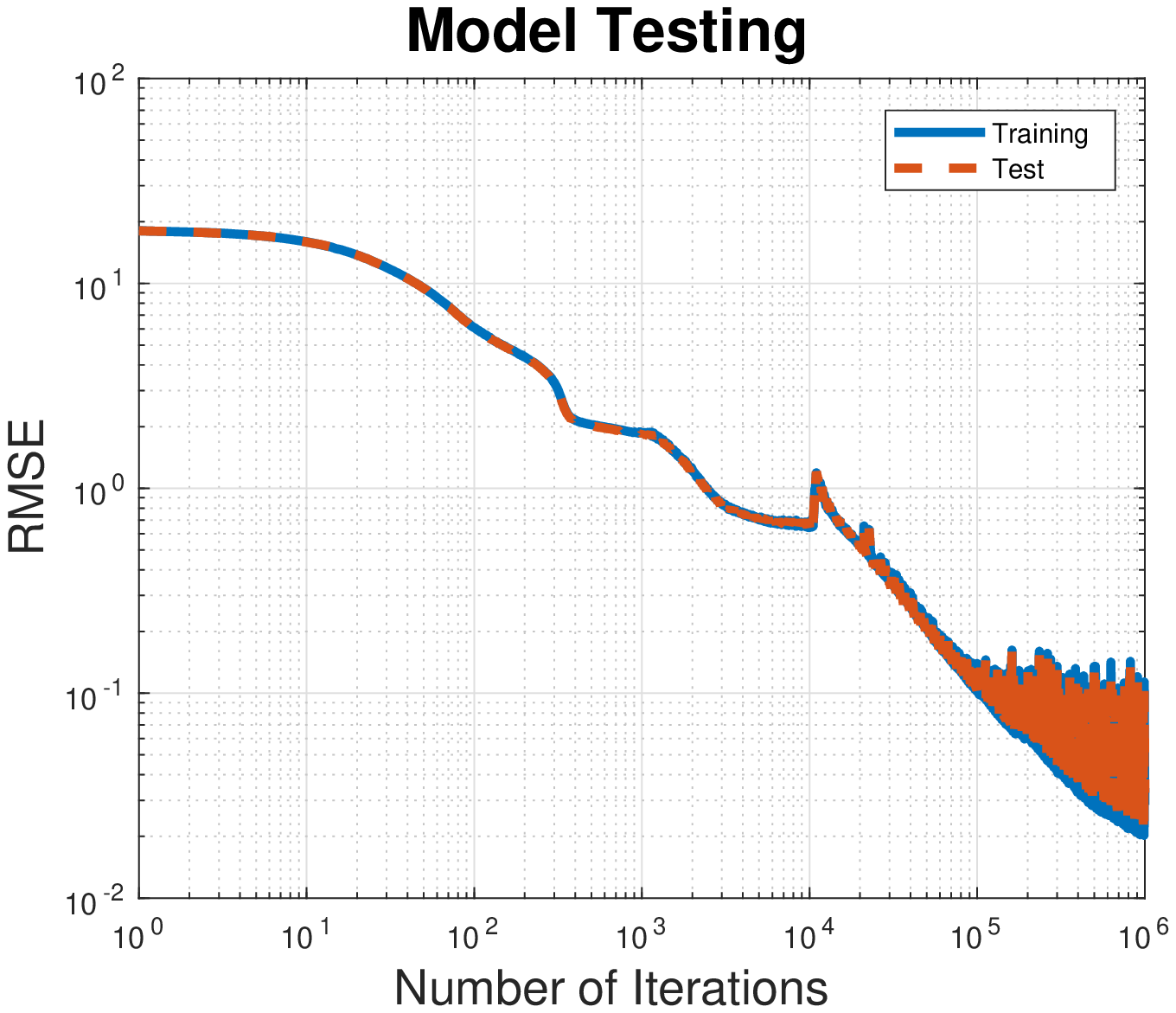}
    \caption{}
    \label{fig:testing}
  \end{subfigure}
  \caption{Model validation and testing}
  \label{fig:validation+training}
\end{figure}

The results for the GAN-MRF model training, validation, and testing
are shown in Fig.~\ref{fig:validation+training}. Specifically,
Fig.~\ref{fig:validation} shows how the choice of the regularization
hyperparameter $\lambda$ in the minimax problem~\eqref{eq:minimax_mod} can be
determined by performing a model validation process. The training and
validation root mean square errors with respect to
different choice of the value of the hyperparameter $\lambda$ that
controls the regularization are plotted, where the blue solid curve
represents the training error and the orange dashed curve represents
the validation error. As shown in the plot, both the training and validation errors
are large as the value of $\lambda$ gets too small or too big. The
training and validation errors are both small at $\lambda = 1$ and
$\lambda = 100$. We choose in our model $\lambda$ to be 100 since it
provides better convergence. Fig.~\ref{fig:testing} shows the
performance of the trained model on the training and test sets as the
number of iterations increases in logarithmic scale. The blue solid
curve represents the training error and the orange dashed curve
represents the test error. One can see a
clear decay in both the training and test errors, which indicates that
there's no evidence of high bias, or model under-fitting problem. In addition, the test
error stays closely with the training error, indicating no evidence of
high variance, or model over-fitting problems.

\begin{figure}[h!]
  \centering
  \includegraphics[scale=.55]{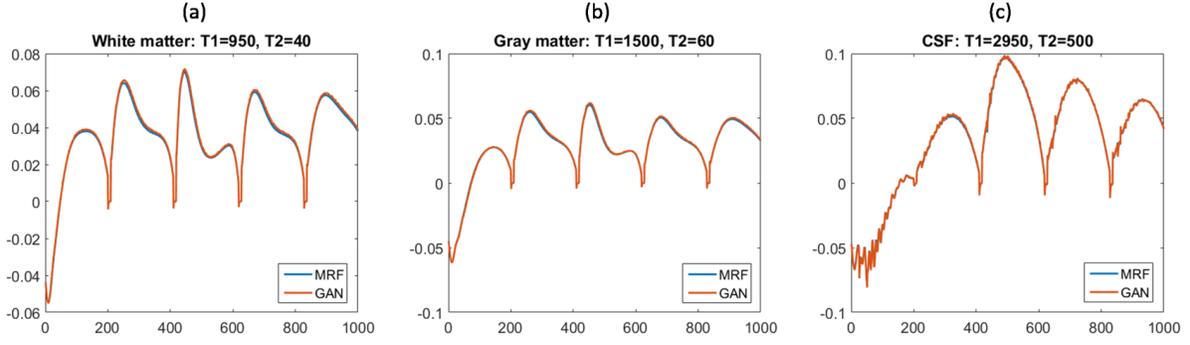}
  \caption{Sample fingerprints generated from GAN-MRF.}
  \label{fig:sample5970}
\end{figure}

Fig.~\ref{fig:sample5970} shows several sample fingerprints generated
by the proposed trained GAN-MRF model, where the orange curves
represent the synthesized fingerprints using the GAN-MRF model, and
the blue curves represent the benchmark MRF fingerprints simulated from Bloch
equations. Fig.~\ref{fig:sample5970}(a) plots a synthesized sample
white matter GAN-MRF fingerprint with $T_1=950$ms and $T_2=40$ms, and
compares it against the corresponding benchmark MRF fingerprint
generated by Bloch simulations. Fig.~\ref{fig:sample5970}(b) shows a
sample gray matter fingerprint generated by the GAN-MRF model with
$T_1=1500$ms and $T_2=60$ms and the corresponding gray matter
benchmark MRF fingerprint. Fig.~\ref{fig:sample5970}(c) shows a sample
CSF fingerprint generated by the GAN-MRF model with $T_1=2950$ms and
$T_2=500$ms and the corresponding CSF benchmark MRF fingerprint. Note that all these GAN-MRF fingerprints match to the MRF-FISP fingerprints well.

\begin{figure}[h!]
  \centering
  \includegraphics{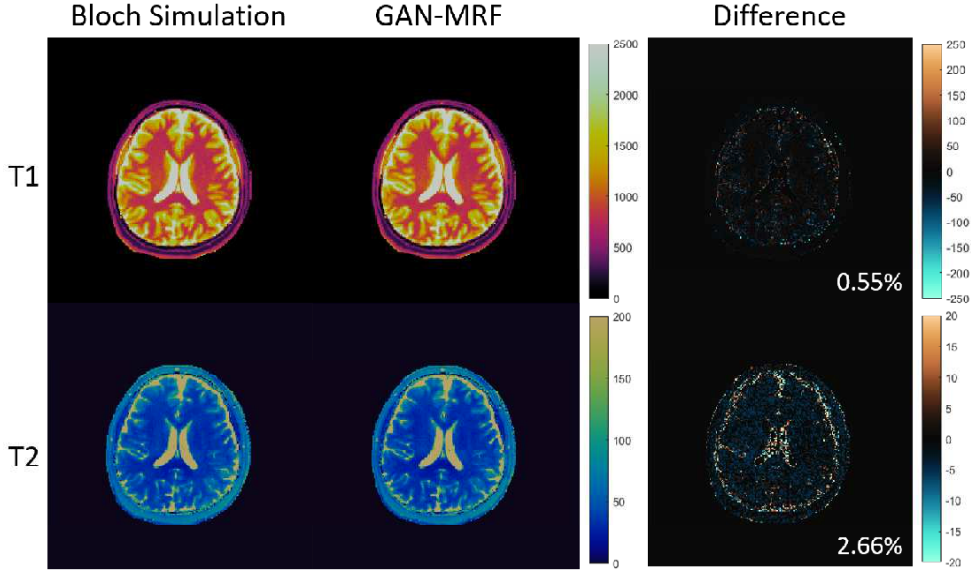}
  \caption{Comparison of maps generated from the MRF-FISP dictionary
    and the GAN-MRF dictionary.}
  \label{fig:result5970}
\end{figure}

We further show the performance of the GAN-MRF model by comparing the
in vivo $T_1$ and $T_2$ maps obtained using the GAN-MRF dictionary
with the benchmark $T_1$ and $T_2$ maps obtained from the MRF-FISP dictionary generated by Bloch
simulations. Shown in Fig.~\ref{fig:result5970}, column (a) are the
benchmark $T_1$ and $T_2$ maps obtained by matching the collected in vivo data
to the MRF-FISP dictionary. Fig.~\ref{fig:result5970} column (b) shows the $T_1$
and $T_2$ maps obtained by matching the collected in vivo data to the
dictionary synthesized by the GAN-MRF model. They show no clear visual
degradation from the benchmark $T_1$ and $T_2$ maps from the first column. The
difference maps scaled 10 times are shown in Fig.~\ref{fig:result5970}
column (c). The relative root mean square error is only 0.55\% for the
$T_1$ maps, and 2.66\% for the $T_2$ maps, which further confirm the finding.

\begin{figure}[h!]
  \centering
  \includegraphics[scale=.6]{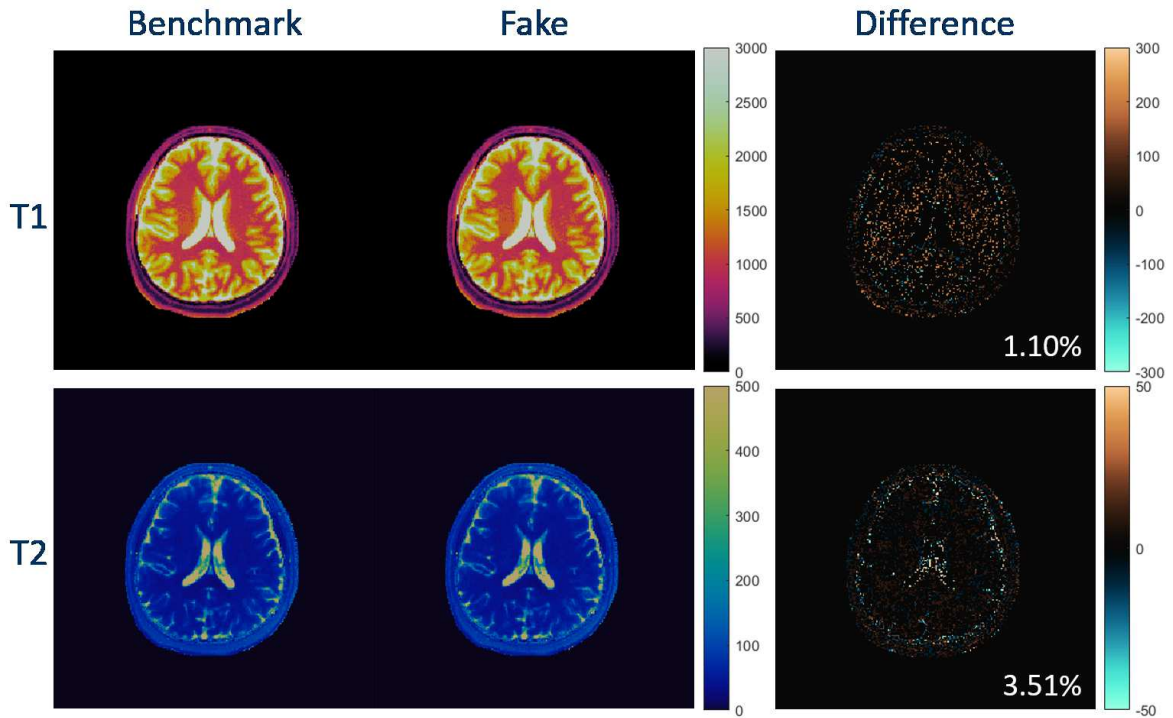}
  \caption{Comparison of maps generated from the coarse MRF-FISP dictionary
    and the GAN-MRF dictionary.}
  \label{fig:result297}
\end{figure}

\begin{figure}[h!]
  \centering
  \includegraphics[scale=.6]{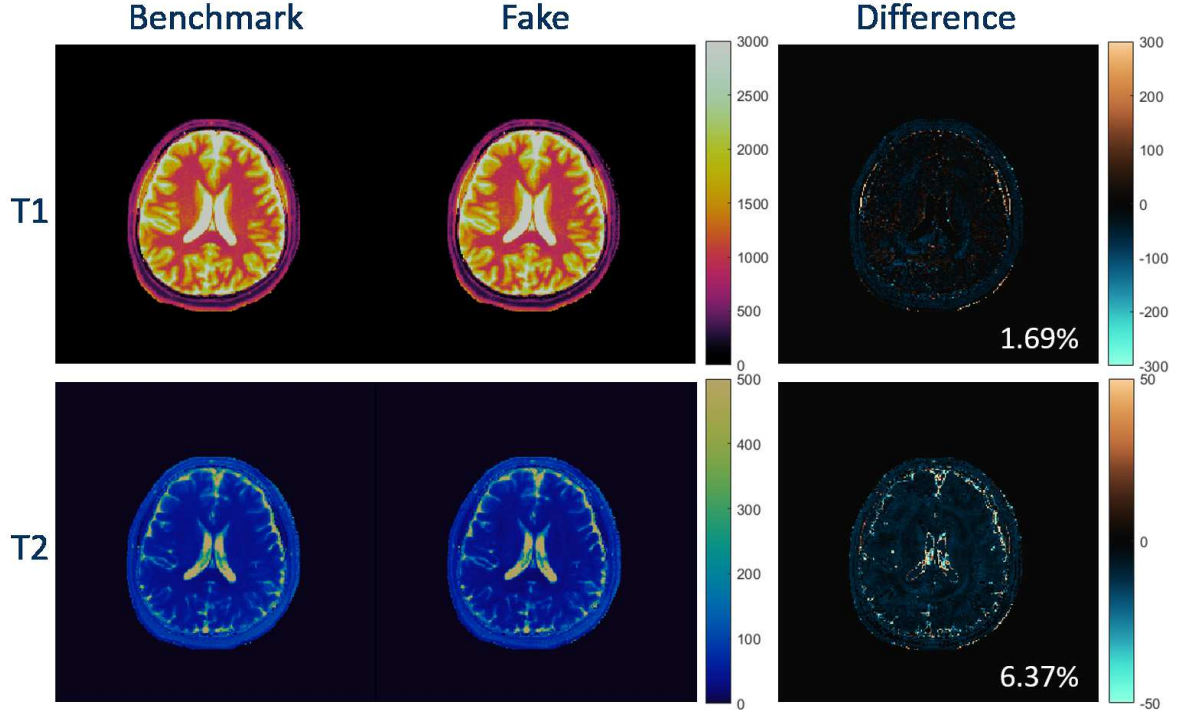}
  \caption{Comparison of maps generated from the fine MRF-FISP dictionary
    and the GAN-MRF dictionary.}
  \label{fig:result106160}
\end{figure}

Fig.~\ref{fig:result297} and Fig.~\ref{fig:result106160} demonstrate
the scalability of our GAN-MRF model. Specifically,
Fig.~\ref{fig:result297} shows the in vivo $T_1$, $T_2$ maps, together with
the corresponding difference maps, obtained from
MRF dictionaries generated from 297 $T_1$, $T_2$ combinations using
the benchmark Bloch simulations and the trained GAN-MRF model with the
same input sequence and tissue parameters respectively. One observes
that,  as expected, the GAN-MRF model works well on this coarse
dictionary. The relative root mean square error for $T_1$ is 1.10\%
and for $T_2$ is 3.51\%. Next, Fig.~\ref{fig:result106160} illustrates the
robustness of the GAN-MRF model. Note that the model is trained on
297 $T_1$ and $T_2$ combinations. The trained GAN-MRF model is then used to
synthesize a dictionary corresponding to 106160 different $T_1$, $T_2$
combinations. The benchmark MRF-FISP dictionary is simulated from the
Bloch equations using the same sequence parameters and the 106160
$T_1$, $T_2$ combinations. Column (a) shows the in vivo $T_1$, $T_2$
maps generated from the benchmark MRF-FISP dictionary via pattern
matching. Column (b) shows the $T_1$ and $T_2$ maps generated from the
synthesized MRF dictionary using the coarsely trained MRF-GAN
model. The difference maps are shown in column (c). We observe from
these figures that by applying the GAN-MRF model trained on the coarse
dictionary, we are still able to get decent $T_1$, $T_2$ maps compared to
those obtained directly from the benchmark MRF-FISP dictionary with the
same number of $T_1$, $T_2$ combinations. The relative root mean
square error for $T_1$ is now 1.69\% and for $T_2$ is 6.37\%, which illustrate good interpolation ability of the trained GAN-MRF model.

Most importantly for the goals here, we would like to see what 
advantage the GAN-MRF model can bring in terms of time consumption for
MRF dictionary generation. As discussed early, the generation of a
slice profile corrected MRF-FISP dictionary requires the simulation of
hundreds of spin evolutions which are then added together for each
time frame to average out the off resonance effect. This process can take
up to hours for a dictionary size of 1000 by 5970 in Matlab depending
on the computer hardware. Note that the part of Bloch equation
simulations is already coded in the C language. Now by using the
GAN-MRF approach, after the GAN-MRF model is trained, it takes only
0.3 second to generate the MRF dictionary with the same size using
Python with the Tensorflow framework. This results in tens of
thousands times speed up. The time needed for training the GAN-MRF
model with the MRF-FISP dictionary of size 1000 by 5970 is 8 hours
by using only the Intel Xeon 2.6GHz CPU.

\section{Discussion}

We have described a deep learning approach for MRF dictionary
generation based on the state-of-the-art generative adversarial
networks. It provides a novel way of mimicking the Bloch equation
simulations from MR physics so as to generate the MRF dictionaries
more efficiently. We have tested, as an example, the FISP sequence
with tissue properties $T_1$, $T_2$ and sequence parameters flip angle
and repetition time as inputs to the model. Our GAN-MRF model reduced
the FISP dictionary generation time from hours to sub-second without
sacrificing much of the performance. Note that in our experiments, we
have only varied the tissue properties $T_1$ and $T_2$. It is also
possible to vary the input flip angles and the repetition
times. Moreover, we see no obstacle to include more sequence and
tissue properties into the GAN-MRF model. Therefore, one should be
able to apply this model to problems with more complicated physics for
rapid calculation of MRF dictionaries. This may increase the time
needed for training, which can be handled with the powerful modern
computer capacity, e.g., the use of GPU computing nodes. Once the
model is trained, it can be used on a basic to mediocre computer to
quickly generate the MRF dictionary. With such a sophisticated deep
learning model, we also believe that it is possible to utilize the
model for optimization problems arising from the end of MRF sequence
parameter design, so that one does not have to tune the sequence
parameters heuristically for optimal performance.

Based on the above discussion, one of the immediate applications of the GAN-MRF model is to the more complicated multi-compartment systems, where complicated models such as MRF-X\cite{Hamilton2015} and EPG-X~\cite{doi:10.1002/mrm.27040} were developed to consider extra properties such as volume fraction, chemical exchange, and magnetization transfer. First, once trained, the GAN-MRF model can help avoid the challenges of generating MRF dictionaries from the complicated models such as the Bloch-McConnell equations in applications, which require specific domain knowledge. Second, a trained GAN-MRF model is much more efficient in generating these MR fingerprints, which can help reduce the turnover time significantly for generating tissue and parameter maps. Moreover, due to the scalability of the GAN-MRF model shown above, the amount of training data needed from these complicated models can be potentially reduced.

Even though our GAN-MRF model has shown great scalability so that one
only needs to train the model using a small set of training data in
some cases, it is possible in other cases that the training MRF
dictionary has a much larger size due to complex physics. In this
case, it is possible to combine the GAN-MRF model with other low rank
approximation methods such as SVD or randomized SVD to reduce the
memory and time consumption for model training. For instance, one may
train the GAN-MRF model with a compressed coarse MRF dictionary using
SVD or randomized SVD. The trained model can then be used to
synthesize a compressed fine MRF dictionary. The tissue and sequence
parameter maps can simply be obtained by applying pattern matching
between the compressed in vivo data and the compressed fine
dictionary. 

The ability of the GAN-MRF model to efficiently mimic the Bloch
equations simulated signals has also great implication for clinical
applications using MR fingerprinting. For instance, in cardiac MR
imaging using MRF, one of the barriers for getting real-time tissue
and sequence parameter maps is the inability of standard methods to
generate a patient specific MRF dictionary on-the-fly, since different
subjects have different heart rates. Therefore, there is no universal
MRF dictionary for cardiac scanning. The proposed GAN-MRF model, on
the other hand, can synthesize Bloch equations simulated signals in
real-time, providing the possibility to generate different MRF
dictionaries according to different heart rates on-the-fly by varying
the repetition time input of the model. 

Last but not least, the scalability of the GAN-MRF model in this paper
is mainly tested on the $T_1$, $T_2$ combinations with the same range except
for Fig.~\ref{fig:result297} and Fig.~\ref{fig:result106160}, where
the starting range of both $T_1$ and $T_2$ for
the coarse and fine dictionaries are different. This together with the
extrapolation ability of the GAN-MRF model need to be further
investigated.

\section{Conclusions}

This work proposed a new approach for MRF dictionary generation based
on the recent development in unsupervised learning, namely, the
generative adversarial networks (GAN). By comparing to the Bloch
equations simulated MRF-FISP fingerprints and the matched $T_1$, $T_2$ maps,
we showed that the proposed GAN-MRF model can generate accurate MRF
fingerprints and as a result, accurate $T_1$, $T_2$ maps with much less
computational time. We further demonstrated that this approach is
robust enough to generate accurate fine MRF maps using the GAN-MRF
model trained from a coarse dictionary. This makes it feasible to
generate on-the-fly new MRF fingerprints with tissue property of
interest as needed. Moreover, it provides the possibility to
significantly reduce the memory and time cost for large scale MRF
dictionary generation for more complicated sequence models. It also
has great potential for real-time MRF mapping in clinical
applications. Furthermore, it opens the door for MRF sequence
parameter optimization problems using deep learning techniques.

\section{Acknowledgments}

The authors would like to acknowledge funding from Siemens Healthcare,
and NIH grants 1R01EB016728-01A1, 5R01EB017219-02.



\bibliography{mri}

\end{document}